\documentclass[bst files/sn-nature,Numbered]{sn-jnl}% Style for submissions to Nature Portfolio journals
\usepackage[utf8]{inputenc}
\DeclareUnicodeCharacter{2009}{\,}

\usepackage{graphicx}%
\usepackage{multirow}%
\usepackage{amsmath,amssymb,amsfonts}%
\usepackage{amsthm}%
\usepackage{mathrsfs}%
\usepackage[title]{appendix}%
\usepackage{xcolor}%
\usepackage{textcomp}%
\usepackage{manyfoot}%
\usepackage{booktabs}%
\usepackage{algorithm}%
\usepackage{algorithmicx}%
\usepackage{algpseudocode}%
\usepackage{listings}%
\usepackage{placeins}
\usepackage{siunitx}

\DeclareSIUnit\fps{fps}
\DeclareSIUnit\cells{cells}

\newcommand{\beginsupplement}{
        \setcounter{table}{0}
        \renewcommand{\thetable}{S\arabic{table}}%
        \setcounter{figure}{0}
        \renewcommand{\thefigure}{S\arabic{figure}}%
     }

\begin{document}

%\title[Article Title]{Bacterial collective motion is scale-free}
\title[Article Title]{Bacterial turbulence  is scale-free}


\author[1]{\fnm{Benjamin} \sur{Perez-Estay}}\email{benjamin.perez-estay@espci.fr}
\author[2]{\fnm{Vincent} \sur{Martinez}}
\author[3]{\fnm{Carine} \sur{Douarche}}
\author[2]{\fnm{Jana} \sur{Schwarz-Linek}}

\author[2]{\fnm{Jochen} \sur{Arlt}}
\author[1,3]{\fnm{Pierre-Henri} \sur{Delville}}
\author[4]{\fnm{Gail} \sur{McConnell}}
\author[2]{\fnm{Wilson C. K.} \sur{Poon}}
\author[1,5]{\fnm{Anke} \sur{Lindner}}\email{anke.lindner@espci.psl.eu}
\author[1,5]{\fnm{Eric} \sur{Clement}}\email{eric.clement@upmc.fr}

\affil[1]{\orgdiv{{Physique et Mécanique des Milieux Hétérogènes (PMMH)}, \orgname{ESPCI Paris, Université PSL, Université Paris Cité, Sorbonne Université, CNRS}, \orgaddress{\street{7}, quai Saint-Bernard, \city{Paris}, \postcode{75005},  \country{France}}}}

\affil[2]{\orgdiv{School of Physics \& Astronomy}, \orgname{The University of Edinburgh}, \orgaddress{\street{James Clerk Maxwell Building, Peter Guthrie Tait Road}, \city{Edinburgh}, \postcode{EH9 3FD}, \country{United Kingdom}}}
\affil[3]{\orgdiv{Laboratoire FAST}, \orgname{Univ. Paris-Sud, CNRS, Universit\'e Paris-Saclay}, \postcode{F-91405}, \city{Orsay}, \country{France}}

\affil[4]{\orgdiv{Strathclyde Institute of Pharmacy and Biomedical Sciences}, \orgname{University of Strathclyde}, \orgaddress{\street{161 Cathedral Street}, \city{Glasgow}, \postcode{G4 0RE}, \state{State}, \country{United Kingdom}}}

\affil[5]{\orgdiv{Institut Universitaire de France (IUF)}, \orgaddress{ \city{Paris}, \country{France}}}


\abstract{

Suspensions of swimming bacteria interact hydrodynamically over long ranges, organizing themselves into collective states that drive large-scale chaotic flows, often referred to as "bacterial turbulence". Despite extensive experimental and theoretical work, it remains unclear whether an intrinsic length scale underlies the observed patterns. To shed light on the mechanism driving active turbulence, we investigate the emergence of large-scale flows in E. coli suspensions confined within cylindrical chambers, systematically varying confinement height over more than two orders of magnitude. We first demonstrate that the critical density for the onset of collective motion scales inversely with this confinement height without saturation, even for the smallest densities observed. Near the onset, both the observed length and time scales increase sharply, with the length scale bounded only by the vertical confinement. Importantly, both scales exhibit clear power-law dependence on the confinement height, demonstrating the absence of an intrinsic length scale in bacterial collective motion. This holds up to scales nearly 10,000 times the size of a single bacterium, as evidenced by transient coherent vortices spanning the full chamber width near the onset. Our experimental results demonstrating that bacterial turbulence is scale-free provide important constraints for theories aiming to capture the dynamics of wet active matter.}

\keywords{wet active matter, collective motion, instability, vortices}

\maketitle
Microscopic living systems, such as sperm cells, bacteria, or epithelial cells exhibit striking examples of collective motion \cite{rothschild_measurement_1949,poujade_collective_2007,dombrowski_self-concentration_2004,zhang_collective_2010},
sharing many similarities with the dynamical organization of more complex living organisms, including flocks of birds or human crowds \cite{vicsek_collective_2012}. 
 %A notable example of collective organization is the emergence of ``turbulent-like'' flow patterns induced by motile bacteria. This phenomenon has become a key case study in the field of out-of-equilibrium physics and active matter \cite{marchetti_hydrodynamics_2013}. 
Among them, bacterial suspensions are especially notable for giving rise to turbulent-like flow patterns, including vortices and jets, consistently observed across various species of swimming bacteria and under different experimental conditions \cite{alert_active_2022}. The observed coherent flow structures are the result of the forces and torques exerted by the bacteria on the surrounding fluid.  Such bacterial suspensions are often considered the epitome of "wet active matter" and significant efforts have been made to establish suitable hydrodynamic models \cite{saintillan_active_2013,marchetti_hydrodynamics_2013} to describe the observations.

However, contradictory evidence and claims emerge from various experiments and theoretical approaches, in particular with regard to the selection of the lengthscale underlying the observed patterns. Several experimental studies claim that the size of the patterns increases with confinement height and bacterial density, until it saturates at a mesoscopic length scale. This scale may vary between studies, but is typically found to be on the order of tens to hundreds of microns, depending on the bacterial species and specific experimental conditions \cite{wensink_meso-scale_2012,sokolov_physical_2012,wioland_directed_2016, guo_symmetric_2018}. Motivated by the interpretation of the saturated length scale as intrinsic, phenomenological models have incorporated explicit length scale selection to reflect this observation \cite{dunkel_fluid_2013,alert_active_2022}. However, recent experiments reported the absence of saturation and a square root dependence on confinement at least up to \qty{400}{\micro\meter}, raising questions about the genuine existence of a mesoscopic lengthscale \cite{wei_scaling_2024}. 
The assumption of an intrinsic length scale also starkly contrasts with predictions from kinetic theory models of bacterial collective motion, which treat each swimming bacterium as an elongated pusher like swimmer, modeled by a force dipole, generating long-range interactions within the entire suspension \cite{saintillan_active_2013, lauga_hydrodynamics_2009}. These models predict that a suspension of randomly oriented bacteria is linearly unstable at long-wavelengths, leading to the spontaneous emergence of bacteria alignment and subsequently collective motion \cite{saintillan_instabilities_2008}. The critical density for instability onset depends on the level of noise in the bacterial orientation and the sample confinement \cite{martinez_combined_2020, hohenegger_stability_2010}. The minimal critical density is reached in the limit of unconfined suspensions, where it is solely given by the bacterial reorientation process \cite{subramanian_critical_2009,stenhammar_role_2017}. The aligned state is itself linearly unstable to bending instabilities and together with additional effects such as boundary interactions they drive the dynamics towards turbulence like patterns in the nonlinear regime \cite{ramaswamy_active-filament_2007, saintillan_active_2013}. Together, these kinetic models consistently predict that the dominant flow structures emerge at the largest accessible scale, which indicates the absence of an intrinsic lengthscale \cite{theillard_geometric_2017, theillard_computational_2019, bardfalvy_particle-resolved_2019, palmer_correlations_2025}. 
%
%The apparent contradiction with experimental observations of a fixed scale is often attributed to short-range interactions or direct alignment mechanisms \cite{sokolov_physical_2012, heidenreich_hydrodynamic_2016, palmer_correlations_2025}.

%%%%%%%%%%%%%%%%%%%%%%%%%%%%%%%%%%%%%%%%%%%%
To answer the fundamental question of the size selection of turbulent flow patterns in bacterial suspensions, with the objective to test the potential existence of an intrinsic length-scale, we present a comprehensive experimental study on semi-dilute suspensions of \textit{E. coli} bacteria (smooth swimmers and wild-type) of volume fraction $\phi$, confined within a cylindrical cavity of height $H$ and radius $R = \qty{5}{\milli\meter} > H$, bounded by no-slip surfaces (Fig.~\ref{fig_transition} a). We monitor the motion of passive fluorescent beads at the midplane, capturing the collectively induced flows (Fig. \ref{fig_transition} b). Using all surfaces permeable to oxygen allows us to maintain bacterial activity for large densities and high confinement heights. By performing 140 distinct experiments with different pairs of $\phi$ and $H$, we determine the critical density $\phi_c$ for the onset of the chaotic flows as a function of the confinement height $H$, showing quantitative agreement with kinetic theory. This allows us to express the selected length and time scales for every experiment as a function of their relative density $\phi / \phi_c$ and the confinement height $H$. We observe that the length scales increase nearly linearly with  $H$. Moreover, near the transition, we observe the possibility of large coherent flows set by the radius $R$, the largest scale in the system.

\subsection*{Onset of bacterial collective motion}

\begin{figure}[tp]%
\centering
\includegraphics[width=\textwidth]{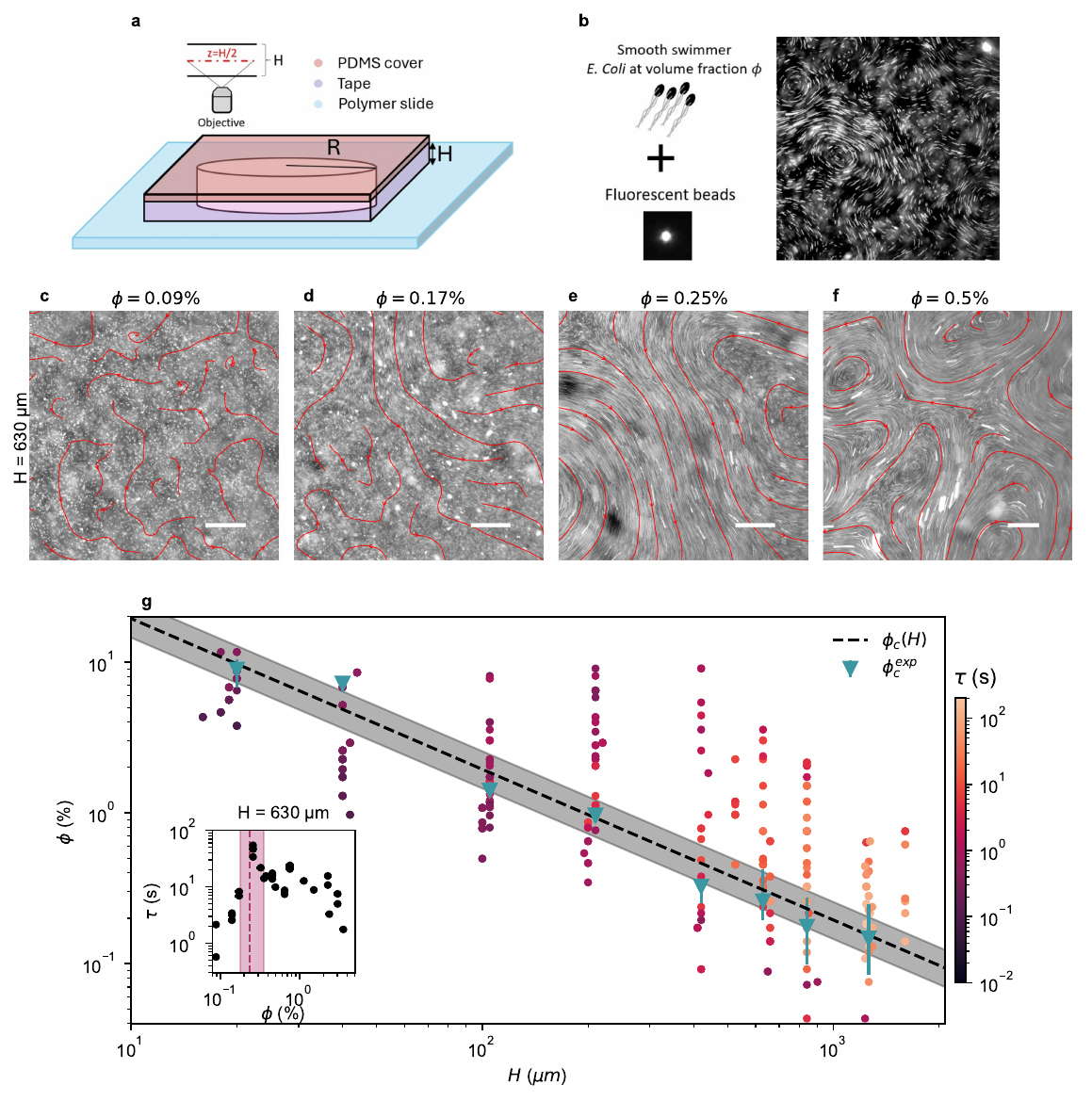}
\caption{\textbf{Critical density for the onset of chaotic flows in bacterial suspensions is inversely proportional to the confinement height $H$.} 
\textbf{a} Sketch of the experimental set-up showing the cylindrical pool of radius $R$ confining the bacterial suspension via two horizontal gas-permeable slides separated by a height $H$. The visualization plane is at mid-distance between the slides.
\textbf{b} The bacterial suspension is composed of bacteria at volume fraction $\phi$ and passive fluorescent beads.
\textbf{c-f} Streaklines showing trajectories of the passive fluorescent beads (60 images at 20 FPS) together with instantaneous streamlines obtained from PIV measurement shown in red. Images correspond to different bacterial densities $\phi$ below (c-d) and above (e-f) the critical density, for the same height $H=$~\qty{630}{\micro\meter}. Scale bars are \qty{100}{\micro\meter}.
\textbf{g} Phase diagram in the experimental parameter space $\phi, H$. Each dot corresponds to one experiment, with a fixed radius of the cylindrical pool of $R=$~\qty{5}{\milli\meter}. The cyan triangles correspond to the experimentally obtained values of the critical density $\phi_c^{exp}$. The black dashed line represents a fit of the kinetic theory prediction $\phi_c = l_{exp}H^{-1}$ \cite{hohenegger_stability_2010} to the experimentally obtained values. The fitting constant $l_{exp} = $~\qty{1.7}{\micro\meter} ($95\%$ CI $[1.4, 2.1]$~\qty{}{\micro\meter}) is in good agreement with the estimated theoretical value $l_{th} \in [1, 3.5] $~\qty{}{\micro\meter} (Supplemental Material). Inset shows the velocity correlation time scale $\tau$ (Methods) for all experiments with $H=$~\qty{630}{\micro\meter}. The maximum of $\tau$ with respect to $\phi$ defines the critical density $\phi_c^{exp}$ for a given height, shown as the purple dashed line. The colored area represents the $95\%$ confidence interval for $\phi_c^{exp}$ (Methods).
}
\label{fig_transition}
\end{figure}

Figure \ref{fig_transition} c-f shows the general behavior of bacterial suspensions when the volume fraction is increased at a constant confinement height. At low concentrations (c-d), a more erratic motion of the tracer beads is observed. As the concentration increases the tracer dynamics evolve into large-scale chaotic vortices and jets  (e-f). Although the flow pattern becomes clearly defined only at higher concentrations, it already exhibits some structure even at low concentrations.  To determine the onset of collective motion at a critical density $\phi_c(H)$ quantitatively, we consequently use an alternative criterion. We measure the flow velocity fields and compute the velocity time-autocorrelation function $C(t)$, with its decay at time scale $\tau$, characterizing the flow dynamics (Methods). Kinetic theory \cite{skultety_swimming_2020} and agent based numerical simulations of pusher type swimmers  \cite{bardfalvy_particle-resolved_2019} have shown that $\tau$ peaks at $\phi_c$, which is consistent with our experimental results. Starting from the lowest densities, we observe an abrupt, two-orders-of-magnitude increase in $\tau$, indicative of a critical slowing down of the dynamics (inset Fig.~\ref{fig_transition}g). We associate the maximum of $\tau$ with the critical density and use this criterion to experimentally determine $\phi_c^{exp}$ as a function of the confinement height $H$ (cyan triangles in Fig.\ref{fig_transition} g, Methods). This approach represents the first experimental determination of the critical density $\phi_c$ through the dynamical criterion of critical slowing down, offering a novel and quantitative method to identify the onset of collective motion in bacterial suspensions.

The experimental values $\phi_c^{exp}$ for smooth swimmers satisfy a scaling relation $\phi_c = l_{exp} H^{-1}$ (black dashed line in Fig.~\ref{fig_transition} g) over the entire set of experiments, spanning confinement heights from \qty{20}{\micro\meter} to \qty{1600}{\micro\meter} and leading to critical densities from approximately $9\%$ to $0.14\%$. Surprisingly, despite the presence of rotational noise in the bacterial suspension, this scaling is in quantitative agreement with the theoretical predictions from kinetic theory obtained for suspensions of dipolar swimmers in the absence of noise-induced reorientation \cite{hohenegger_stability_2010}, with the experimentally determined proportionality constant $l_{exp} = \qty{1.7}{\micro\meter}$ (95\% CI 1.4-2.1~\qty{}{\micro\meter}) matching closely the theoretically predicted value $l_{th} \in [1.7, 3.5]$~\qty{}{\micro\meter} \cite{hohenegger_stability_2010} (Supplemental Material). In the limit of an unconfined suspension, the critical density has been predicted to saturate at a constant value $\phi_c(H\to \infty)$ given by the level of orientational noise in the system \cite{subramanian_critical_2009}. In contrast, in our experiments the linear scaling with the inverse of the confinement height is observed to hold even for the largest heights, reaching critical densities below the predicted saturation values (Supplemental Material).

Moreover, we observe large collective flows in wild-type tumbling bacteria at similar densities, despite the expected strong effect of tumbling on bacterial reorientation, suggesting that the relation between random bacterial reorientation and $\phi_c(H\to \infty)$ is still not well understood (Extended Data Fig.~\ref{sfig_RT_results}).

At the smallest heights we observe critical densities of around $9\%$ approaching the regime where short-range interactions or direct alignment mechanisms are expected to become important \cite{sokolov_physical_2012, heidenreich_hydrodynamic_2016, palmer_correlations_2025} most likely limiting the extension of the scaling to even smaller confinement heights.

The robustness of the scaling $\phi_c\propto H^{-1}$ across a wide range of confinements provides $\phi / \phi_c$ as a reliable parameter to control the distance to the transition for experiments with different heights. 

\subsection*{Scaling of flow structures with confinement height and density}

\begin{figure}[!t]%
\centering
\includegraphics[width=\textwidth]{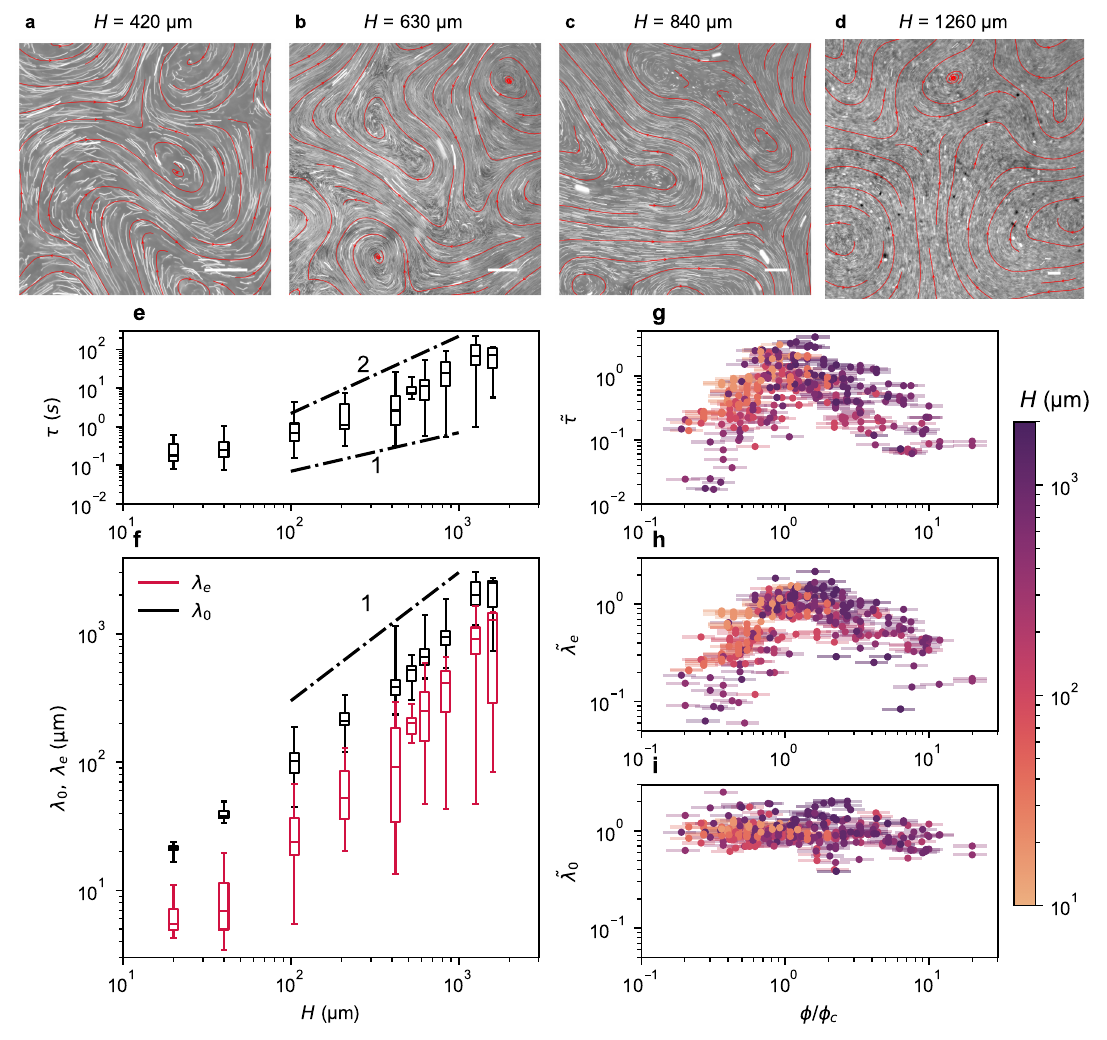}
\caption{\textbf{Confinement $H$ defines the time and length scales of bacterial collective motion.}
\textbf{a-d} Streaklines showing trajectories of the passive fluorescent beads advected by the collectively induced flows (60 images at 20 FPS) together with instantaneous streamlines obtained from PIV measurement shown in red. The observation windows are rescaled to three times the height $H$. Scale bar is \qty{100}{\micro\meter}. 
\textbf{e-f} Box plots for the characteristic time and length scales as a function of confinement height $H$, including all experimental data. The box represents the interquartile range (contains $50\%$ of the data), while the line inside marks the median. The thin whiskers indicate the entire range of data. 
\textbf{g-i} Normalized scales $\tilde{\tau}$, $\tilde{\lambda}_e$, and $\tilde{\lambda}_0$ versus the relative density $\phi/\phi_c$ (see text). Experiments are color-coded with the height and error bars corresponding to the estimated error of $\phi/\phi_c$. 
}
\label{fig_scaling} 
\end{figure}

Having determined the relative density $\phi / \phi_c$ with respect to the density at the transition, we now turn to the scaling of the patterns with respect to the height $H$ and ultimately their overall dependence with respect to $\phi$ and $H$ combined. In Figure \ref{fig_scaling} a-d, we present streamlines for $\phi\approx 2 \phi_c$ and confinement heights $H>$~\qty{420}{\micro\meter} with vortical structures noticeably larger than previously reported \cite{martinez_combined_2020,wei_scaling_2024}. Visual inspection of Figures \ref{fig_scaling} a-d shows that the characteristic size of the flow structures increases with $H$. We characterize the length scale of the patterns by the spatial correlation function $C(r)$, with its decay length $\lambda_e$ and the crossover to zero $\lambda_0$ (Methods). $
\lambda_e$ characterizes the short-range decay of $C(r)$, while $\lambda_0$ marks the onset of negative correlations associated with vortices. Indeed, Figure \ref{fig_scaling} e-f shows that both time and length scales are monotonously increasing with $H$, showing no apparent saturation even for the largest confinement heights (up to \qty{1.6}{\milli\meter}). Interestingly, the scaling of length and time scales with confinement height is similar for wild-type bacteria (Extended Data Fig. \ref{sfig_RT_results}).

Although both time and length scales follow a power law with $H$, the box plots for each height cover one or two orders of magnitude. Thus, to elucidate the dependence with respect to the distance from the instability onset, we propose that the scales $\gamma$ ($\gamma$ being $\tau$, $\lambda_e$ or $\lambda_0$) can be written as $\gamma (\phi, H) = \tilde{\gamma}(\phi/\phi_c)(H/h_\gamma)^{\kappa_\gamma}$ with $h_\gamma$ a normalization constant and $\kappa_\gamma$ a scaling exponent. The scaling exponents $\kappa_\tau$ ($\kappa_\tau = 1.36$ (95 \% CI 1.1-1.56), $\kappa_{\lambda_e} = 1.13$ (1.05-1.2) and $\kappa_{\lambda_0} = 1.04$ (1-1.07)) together with $h_\gamma$ are obtained by fitting only data for bacterial densities close to the instability onset and setting $\tilde{\gamma}=1$ (Extended Data Fig.\ref{sfig_scaling_exponents}). These values are then used to represent $\tilde{\gamma}(\phi/\phi_c)$ as a function of $\phi/\phi_c$ for the full range of densities.  Based on this simple scaling ansatz, we obtain roughly a data collapse for all rescaling functions $\tilde{\tau}$, $\tilde{\lambda}_e$, and $\tilde{\lambda}_0$ (Fig.~\ref{fig_scaling} g-i).   Similarly to the correlation time $\tilde{\tau}$, the length scale $\tilde{\lambda}_e$ increases significantly when approaching $\phi / \phi_c = 1$, and is found to slowly decrease for densities above the critical density, again, in agreement with predictions from agent based numerical simulations \cite{bardfalvy_particle-resolved_2019}. This finding highlights the importance of controlling the distance from instability onset when determining scaling laws in bacterial collective motion. On the other hand, the correlation length $\tilde{\lambda}_0$,  is independent of $\phi / \phi_c$ with a normalization constant $h_{\lambda_0} = $~\qty{1.1}{\micro\meter}, meaning $\lambda_0$ on average is essentially equal to the confinement height $H$.

The scatter in the rescaling is attributed not only to experimental uncertainties in $\phi/\phi_c$, but also to intrinsic fluctuations arising from the complex, nonlinear dynamics of this out-of-equilibrium system. Indeed, we observe that the length scales of the flow present large fluctuations in time and do not always remain bounded by $H$ (Extended Data Figure~\ref{sfig:length_fluctuations_alternative}). The particular collective flows we discuss next illustrate this important aspect. 

\subsection*{Emergence of large vortex states}
\begin{figure}[!t]%
\centering
\includegraphics[width=\textwidth]{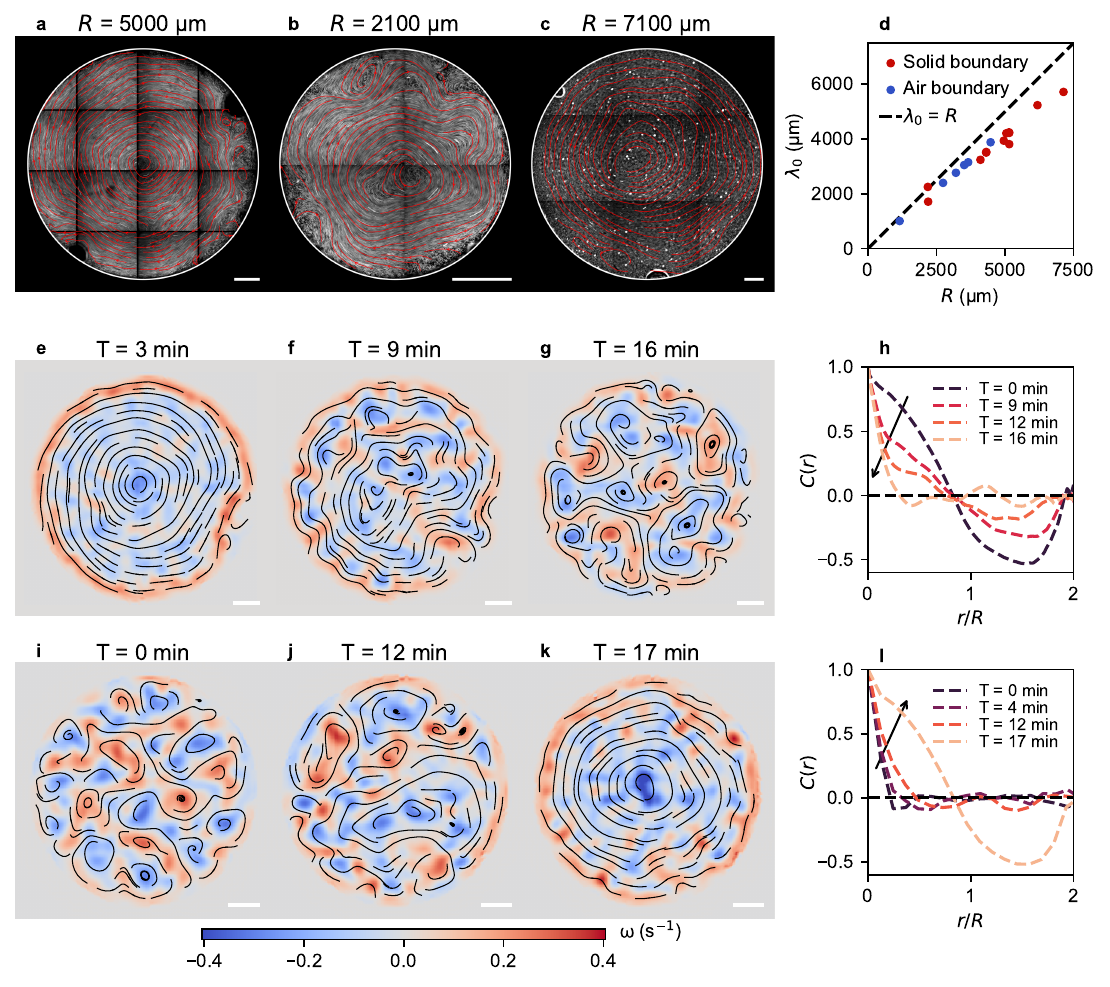}
\caption{\textbf{Single large vortex states (SLV) observed in bacterial suspensions and their breakdown and coarsening dynamics.}
\textbf{a-c} Snapshots of SLV state formed in cylindrical confinement, in the case of no-slip boundaries at the perimeter, indicated by the white line. Images are constructed by superimposing multiple fields of view with \qty{5}{\second} of images each. Titles indicate the radius of the sample $R$. 
\textbf{d} Correlation length scale $\lambda_0$ for different pool radii $R$ for SLV states. Colors indicate the type of boundary condition at the perimeter, with red indicating a solid boundary and blue an air-liquid interface. 
\textbf{e-g} Snapshots of the vorticity field $\omega=|\nabla \times \bf{v}|$ during the breakdown of a SLV for $R=$~\qty{5}{\milli\meter} for solid outer boundary conditions. Initially, the flow corresponds to a single large vortex. In the intermediate snapshot at $t=$~\qty{7}{\minute}, smaller structures appear in the flow, but the core of the vortex is still present. Finally, the smaller structures overcome the large vortex. At this stage, the suspension has fully developed a chaotic flow pattern. 
\textbf{h} Correlation function time evolution for the example shown in \textbf{e-g}. The arrow corresponds to the evolution in time.
\textbf{i-k} Snapshots of a SLV formation for $R=$~\qty{4}{\milli\meter} for solid outer boundary conditions. Initially, the flow is chaotic, and a large flow structure gradually grows in size through the coarsening of smaller vortices.
\textbf{l} Correlation function $C(r)$ for the example shown in \textbf{i-k}. The arrow corresponds to the evolution in time.
Scale bars are \qty{1}{\milli\meter}. 
}
\label{fig_large_state_scaling} 
\end{figure}
Occasionally, we observed very large collective flow structures in the form of a single large vortex (SLV) spanning the entire geometry laterally. Due to the coherence of this large-scale flow it is possible to reconstruct it by visualizing subdomains of the sample (Methods). Figure~\ref{fig_large_state_scaling} a shows an SLV state under similar geometric conditions as the previously discussed experiments ($R=$~\qty{5}{\milli\meter} and solid outer boundary conditions). To further test the conditions under which these states form, we have varied the radius $R$ and the outer boundary conditions (solid or air boundary) showing two other examples in Figures~\ref{fig_large_state_scaling} b-c. Despite the emergence of these states not being a frequent feature across our experiments, we could repeatedly observe their formation for densities between one and three times $\phi_c$ and for aspect ratios $R/H<10$, being more likely to appear for smaller aspect ratios $R/H$. For these flow structures, the length scale $\lambda_0$ equals the largest scale available, the cavity radius $R$ (Fig.~\ref{fig_large_state_scaling} d). It is important to note that at this scale, the Reynolds number of the flow reaches values as large as $Re=0.26$, eventually challenging the Stokes-flow assumption of the theoretical frameworks (Extended Data Fig.~\ref{fig:SLV_Re}). 

The SLV states are transient and exhibit complex dynamics that depend on the boundary conditions and the aspect ratio $(R/H)$. While further details are provided in the Supplemental Material, two representative examples, shown in Figure~\ref{fig_large_state_scaling} (e-h) and (i-l), illustrate key aspects of their evolution. The vortex structure is clearly visible through the representation of the vorticity field $\omega$. In the most common scenario (Fig.~\ref{fig_large_state_scaling}e–h), the SLV is already fully developed at the start of observation $( T = 0 )$ and gradually destabilizes: streamlines bend, smaller vortical structures emerge at the boundaries, and the system evolves into a chaotic state. This transition is accompanied by a shift in the correlation function from a crossing at $\lambda_0 \sim R$ to one at $\lambda_0 \sim H $, reflecting the change in the dominant flow scale. In contrast, the second, less frequent scenario (Fig.~\ref{fig_large_state_scaling}i-l) captures the formation of the SLV from an initially disordered flow composed of small vortices. Through a coarsening process, these vortices merge into a large-scale coherent structure, which persists for tens of minutes before eventually collapsing again into the same chaotic state.

\subsection*{Discussion}

Our experimental study clearly shows that the collective motion spontaneously emerging in a suspension of motile bacteria is a scale-free phenomenon. Over a wide range of confinements and bacterial densities, the chaotic vortices and jets observed scale nearly linearly with the vertical confinement height $H$ (up to \qty{1.6}{\milli\meter}). The observation of transient single large vortex (SLV) states, whose radial extension is determined by the largest lateral confinement $R$ (up to \qty{5}{\milli\meter}), reinforces this point. This result sharply contrasts with the concept of an intrinsic mesoscopic scale often introduced in models to capture the ``active turbulence'' phenomenology \cite{alert_active_2022, wensink_meso-scale_2012,sokolov_physical_2012,wioland_directed_2016, guo_symmetric_2018}. Our finding remains qualitatively consistent with the general expectations of kinetic theory of pusher type elongated swimmers, which predicts a long-wavelength instability for an initial homogeneous and isotropic state \cite{saintillan_active_2013} only limited by the system size \cite{palmer_correlations_2025} as well as the existence of large vortex states in 2D \cite{theillard_geometric_2017}. We demonstrate a two-way dynamical link between the SLV and the chaotic state, which can occur for aspect ratios $R/H$ as large as 10 underscoring the presence of a complex, nonlinear pattern selection mechanism \cite{nishiguchi_vortex_2025} whose underlying principles remain to be fully understood.

We determined the critical density for the onset of collective motion of smooth swimmer suspensions with respect to confinement height, showing an inversely proportional scaling, 
%$\phi_c = l_{exp}H^{-1}$  with the proportionality constant 
in quantitative agreement with predictions of kinetic theory for noiseless pusher type swimmers \cite{hohenegger_stability_2010}. This scaling extends to densities smaller than the limit where Brownian fluctuations are expected to prevent the onset of collective motion \cite{subramanian_critical_2009,koch_collective_2011}, hinting at a dynamical suppression of the random reorientation processes. The negligible role of orientational noise in the onset of collective motion is further supported by the observation of sustained collective motion in suspensions of wild-type bacteria for similarly small densities. 

Around the critical concentration $\phi_c$, we observe a dynamical slowing down and a length scale divergence, limited only by the finite size of the system. This behavior closely resembles the finite-size scaling generically associated with critical, second-order-like transitions \cite{mon_finite_1992,domb_phase_2000} and has recently been predicted from kinetic theory \cite{skultety_swimming_2020} and reported in Lattice Boltzmann agent-based simulations \cite{bardfalvy_particle-resolved_2019}. These observations, in conjunction with previous reports of vanishing viscosity \cite{lopez_turning_2015,martinez_combined_2020}, support the idea that bacterial suspensions can indeed be described as a 'critical' fluid, whose constitutive and transport properties remain to be elucidated.

Overall, our experimental findings shed new light on the scale selection in bacterial turbulence for an important class of ``wet active matter''. They also raise theoretical challenges, particularly in elucidating the role of orientational noise in the onset of collective motion and in identifying the nonlinear mechanisms that govern dynamical pattern selection with respect to the relevant system sizes, which have so far remained unaddressed.

\section*{Methods}

\subsection*{Sample preparation}
\label{methods:sample_prep}

Experiments reported in the main text were performed with the non-chemotactic, smooth swimmer strain of \textit{E.coli} JEK1038 (W3110 [lacZY::GFPmut2, cheY::frt], GFP protein not induced for this study). Some experiments were also done with the chemotactic counterpart JEK1036 (W3110 [lacZY::GFPmut2]). The bacteria were grown in Lysogeny Broth (LB) until the optical density at \SI{600}{\nano\meter} (OD) reached $0.6\pm 0.1$. Then, the bacteria were centrifuged and resuspended in a motility buffer (MB) (10 mM K2HPO4, 10 mM KH2PO4, 10 mM sodium lactate, 0.1 mM EDTA, 0.1 mM L-methionine, 0.2 mM L-serine). MB is a minimal medium that prevents cell division but allows bacteria to swim at typical speeds of $25\pm 5$~\qty{}{\micro\meter\per\second} . A calibration between optical density and bacterial density $n$ was performed using a Malassez counting cell, giving the relation $n = k$OD with $k=(7.7 \pm 0.4) \times 10^8$~\qty{}{\cells\per\milli\liter} (Supplementary Methods). We use the typical bacteria volume $\mathcal{V}_b =$ \qty{1.4}{\cubic\micro\meter} \cite{jepson_enhanced_2013} to compute the bacteria volume fraction as $\phi= \mathcal{V}_b n$. Using the calibration between density and OD, we report bacterial densities in terms of volume fraction $\phi = 1.08 \times 10^{-3}$OD. Passive fluorescent particles (ThermoFischer Cat No. R0300/R300 radius $r_p=$~\qty{3}{\micro\meter} at volume fraction smaller than $ 0.1\%$ depending on the magnification used) were added after measurement of the bacterial density. This final dilution is taken into account in the reported bacterial density. 

Bacterial suspensions were loaded into cylindrical pools of radius R and height H. The resulting flow dynamics was monitored as a function of bacterial density ($10^8$ to $10^{11}$ cells per milliliter) and confinement height H ( \qty{20}{\micro\meter} to \qty{1600}{\micro\meter} ). Meanwhile, the radius was fixed at $R=$~\qty{5}{\milli\meter} for the majority of our experiments, and only changed when studying the formation of SLV states (see main text). For heights of $H =$~\qtylist{20;40}{\micro\meter} we used standard lithography techniques to produce resin molds of rectangular channels with thickness $H=$~\qtylist{20;40}{\micro\meter} and fixed width of $w=$~\qty{600}{\micro\meter}.  For larger $H$, pools were made by stacking double-sided tape (Neschen Gudy 804, thickness $70\pm4$~\qty{}{\micro\meter}) with a circular hole between \qtylist{0.2;1.6}{\centi\meter} diameter attached to a polymer slide (Uncoated: \#1.5 polymer coverslip, hydrophobic, unsterile). When loaded with the appropriate volume of bacterial suspensions, the pools were covered by a PDMS  block \qty{5}{\milli\meter} thick. Using oxygen-permeable surfaces was essential to sustain bacterial activity, even in samples with large heights and high bacterial concentrations. When glass slides were used instead, the lack of oxygen led to a noticeable decrease in bacterial activity within minutes. 
Although we could not directly measure bacterial activity in our experiments, we monitored the collective motion dynamics throughout and discarded any experiments showing a slowdown over time, which guaranteed bacteria remained active.

\subsection*{Image acquisition and analysis}

Images were acquired between \qtylist{5;40}{\fps} at the midplane using an Orca Hamamatsu CMOS camera, filter cube Zeiss 90 HE LED set, and a Zeiss Observer Z1 microscope. With an LED light source (Collibri 7, Zeiss), we excite the sample with a wavelength of  555/\qty{38}{\nano\meter} and the beads emit at \qty{612}{\nano\meter}. Regarding the magnification, it is preferable to have higher magnifications to observe the details of the flow. Nevertheless, the field of view must be at least three times larger than the characteristic pattern size. Therefore, the objective magnification used depends on the specific confinement height and varies between 2.5 and 40 times magnification. 

Image sequences obtained are analyzed using the particle image velocimetry (PIV) algorithm to measure velocities \cite{raffel_particle_2007}. Calculations were performed using PIVlab open-source software \cite{thielicke_particle_2021,stamhuis_pivlab_2014}. In PIV, the image is divided into square interrogation areas that tessellate all space. The cross-correlation matrix with the successive frame is computed for each interrogation area. We used three subsequent interrogation passes with 64, 32, and 32 pixels as the square size, with $50\%$ overlap between squares. Particle seeding is adjusted depending on the magnification, aiming to have about three particles in the last square.

\subsection*{Length and time scales definition}

To define characteristic scales of the collective motion, we use the correlation function of the fluid velocity field, defined as,

\begin{align}
\label{eq:correlation_function}
    C(\textbf{r}, t) &= \frac{\langle \textbf{v}(\textbf{R},T)\cdot\textbf{v}(\textbf{r}+\textbf{R}, t+T) \rangle_{\textbf{R},T} - \langle \textbf{v} \rangle^2 }{\langle |\textbf{v}|^2 \rangle - \langle \textbf{v} \rangle^2 }, \\
    C(t) &= C(0, t), \\
    C(r) &= \langle C(\textbf{r}, 0) \rangle_\theta.
\end{align}

The time autocorrelation function $C(t)$ is the autocorrelation of the velocity field with itself, averaged over all positions $\textbf{R}$ and times $T$. We extract a correlation timescale $\tau$ as the time at which the correlation function reaches $e^{-1}$. The spatial correlation function $C(r)$ averaged over all directions $\theta$ is the correlation between two velocities at a distance $r$. Analogously, we define characteristic lengthscales $\lambda_e$ and $\lambda_0$ as the minimum distances that satisfy $C(\lambda_e)=e^{-1}$ and $C(\lambda_0)=0$ respectively. These two length scales give different information. $ \lambda_e$ is a short length decay, while $\lambda_0$ indicates where the correlation becomes negative. Negative values of $C(r)$ indicate an anti-correlation of the velocities. Vortices are characterized by a rotation of the velocity field, creating areas with velocities pointing in opposite directions. Therefore, the length scale $\lambda_0$ is a proxy for the typical vortex size \cite{wensink_meso-scale_2012}. Seeking to ensure statistical convergence of the correlation functions, the size of the window of observation is always three times the confinement height $H$, and the temporal correlation function is only computed until a fifth of the video length.

\subsection*{Cartography}

To measure the large flow structures observed in the experiments, such as the SLV, we divided the field of view into multiple regions and recorded 5s videos at each position. The imaging process takes approximately 6s per region, including the time required to move the stage. For the largest SLV observed, 16 regions were imaged, resulting in a total acquisition time of about 96 seconds per snapshot. This approach yields a discontinuous measurement of the flow in both space and time. Consequently, it is only applicable when the flow evolves slowly, with correlation timescales on the order of minutes. This technique is therefore constrained to flow regimes with long correlation times, such as those observed in the SLV states.

\backmatter

%\bmhead{Supplementary information}

%\bmhead{Acknowledgments}Acknowledgments are not compulsory. Where included they should be brief. Grant or contribution numbers may be acknowledged. Please refer to Journal-level guidance for any specific requirements.

%\section*{Declarations}Some journals require declarations to be submitted in a standardised format. Please check the Instructions for Authors of the journal to which you are submitting to see if you need to complete this section. If yes, your manuscript must contain the following sections under the heading `Declarations':

\bmhead{Funding} 
B.P-E.,A.L and E.C. acknowledge support of the ANR-22-CE30 grant "Push-pull" and the European program H2020-MSCA-ITN-2020-PHYMOT. All authors thank the CNRS/Royal Society Grant PHC-1576 that started the Paris-Edinburgh collaboration. 

\bmhead{Acknowledgements}
We thank David Saintillan and Brato Chakrabarti for enlightening discussions. We are especially grateful to Alexander Morozov for numerous discussions and his important contributions to shaping this work.   

\bmhead{Conflict of interest/Competing interests} 
The authors declare no conflict of interest.

\bmhead{Data availability}
Data is available on request.

\bmhead{Author contributions}
The research was designed by V.M., W.C-K. P.,C.D, B. P-E, A.L and E.C.. Early experiments have been conducted by C.D, P-H.D, V.M, J.A. J. S-L.  and G McC. All experiments and data curation presented in this paper were performed by B.P-E.  Data analysis and manuscript writing was undertaken by B.P-E, A.L. and E.C.  All authors contributed to the scientific discussions and critically reviewed the manuscript.

%\noindentIf any of the sections are not relevant to your manuscript, please include the heading and write `Not applicable' for that section. 

\begin{appendices}
\section{Extended Data}

\begin{figure}[!htp]%
\centering
\includegraphics[width=\textwidth]{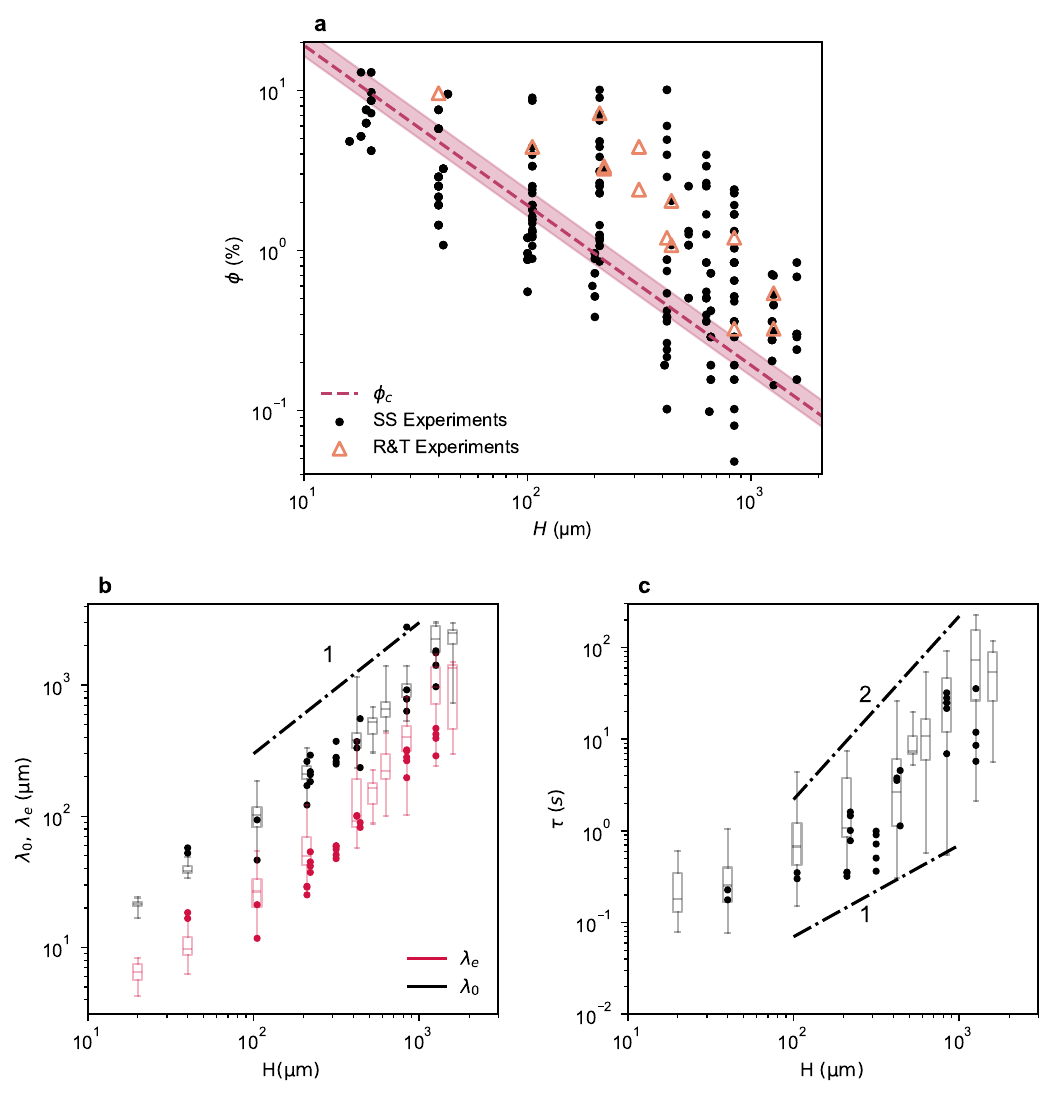}
\caption{\textbf{Run-and-tumble (R\&T) E. coli bacterial suspensions show analogous pattern size scaling to smooth swimming (SS) bacteria.} 
\textbf{a} R\&T bacteria experiments performed represented by orange triangles in the $H,\phi$ parameter space. Black dots correspond to experiments with SS bacteria, shown here as a reference. The purple dashed line represents the fit of the critical density $\phi_c(H) = l_{exp}/H$ (see Figure \ref{fig_transition}). All experiments with R\&T bacteria exhibited collective motion.
\textbf{b} Correlation length scales $\lambda_0$, $\lambda_e$, and \textbf{c} correlation time scale $\tau$ as a function of $H$. Dots represent values obtained for R\&T bacteria, while transparent box plots correspond to experiments with the smooth swimmer strain, shown as reference.}
\label{sfig_RT_results} 
\end{figure}

\begin{figure}[!htp]%
\centering
\includegraphics[width=\textwidth]{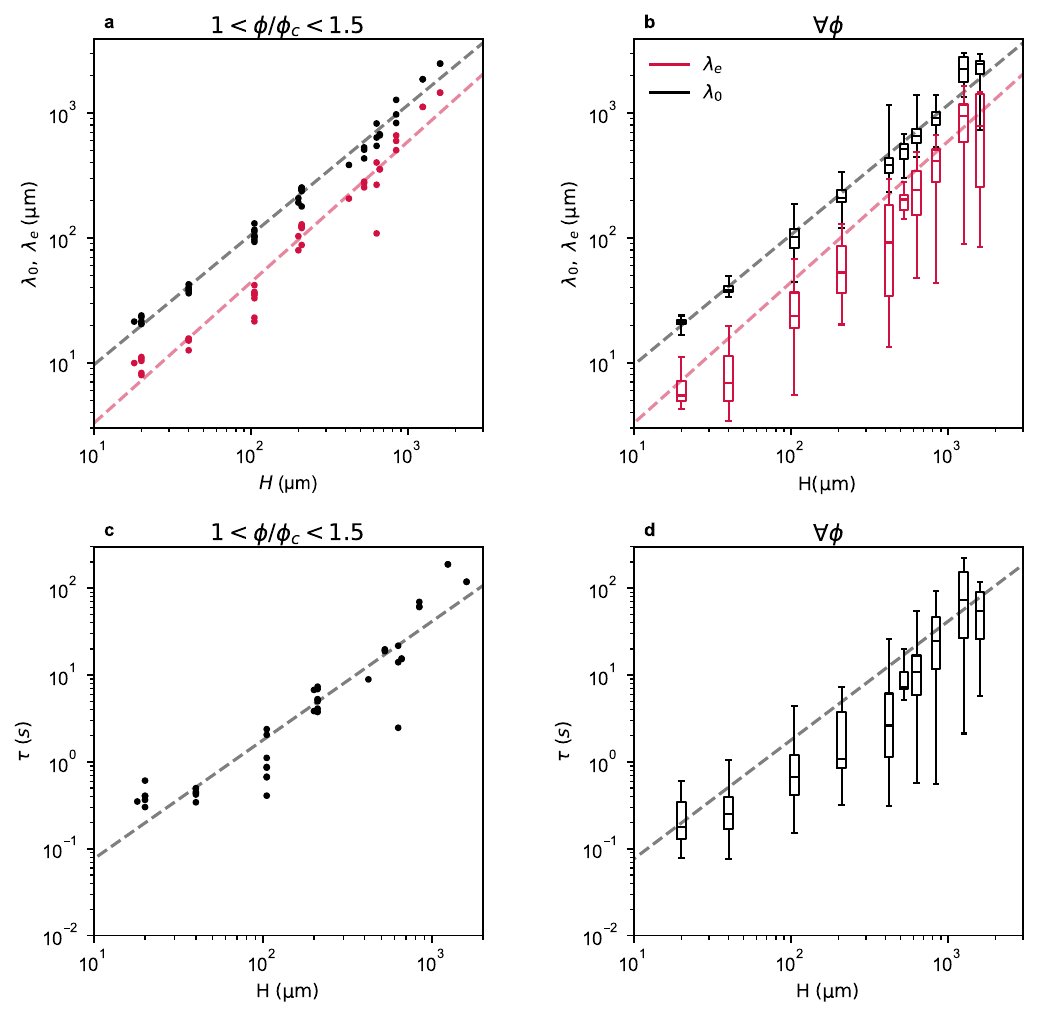}
\caption{\textbf{Scaling exponents for the characteristic length scales.}
\textbf{a} Correlation length scales as a function of the system size $H$, for experiments $1< \phi / \phi_c <1.5$ near the critical transition. The dashed line corresponds to the fit of $\lambda = (H/h_\lambda)^{\kappa_\lambda}$ that represents the scaling of the length scales near the transition. Resulting parameters are $\kappa_{\lambda_e} = 1.13$ ($95\% CI 1.05-1.2$),  $h_{\lambda_e} = 3.5$ ($95\% CI 3-4$), $\kappa_{\lambda_0} = 1.04$ ($95\% CI1-1.7$) and $h_{\lambda_0} = 1.1$ ($95\% CI 1-1.2$). 
\textbf{b} The scaling obtained from \textbf{a} compared to data for all $\phi$ shown as box plots for the distribution of the correlation length scales as a function of $H$.
\textbf{c-d} Analogous figures for the correlation time scale $\tau$. Fitting parameters are $\kappa_\tau = 1.36$ ($95\% CI 1-1.56$) and $h_\tau = 65$ ($95\% CI  55-79$). Confidence intervals of the parameters reported in parentheses were determined from a bootstrap scheme with resampling.
}
\label{sfig_scaling_exponents} 
\end{figure}

\begin{figure}[!htbp]%
\centering
\includegraphics[width=\textwidth]{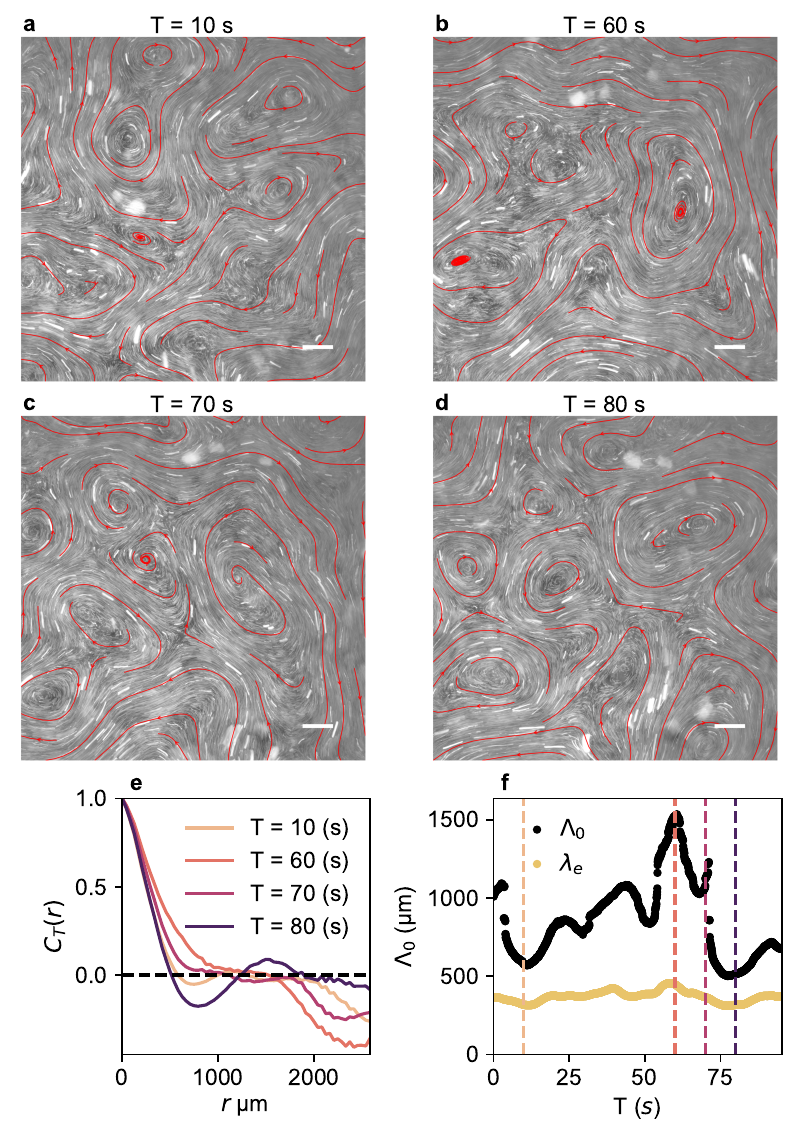}
\caption{ 
\textbf{Example of the time evolution of the characteristic pattern size.}
\textbf{a-d} Four snapshots of patterns observed in a single video. Scale bars are \qty{200}{\micro\meter}. Volume fraction $\phi = 0.5\%$ and height $H = $~\qty{630}{\micro\meter}
\textbf{e} Instantaneous correlation function $C_T$ of the four snapshots.
\textbf{f} Length scales $\lambda_0$ and $\lambda_e$ as function of time $T$. The dashed lines mark the times of the snapshots. 
}
\label{sfig:length_fluctuations_alternative}
\end{figure}

\begin{figure}%
\centering
\includegraphics[width=\textwidth]{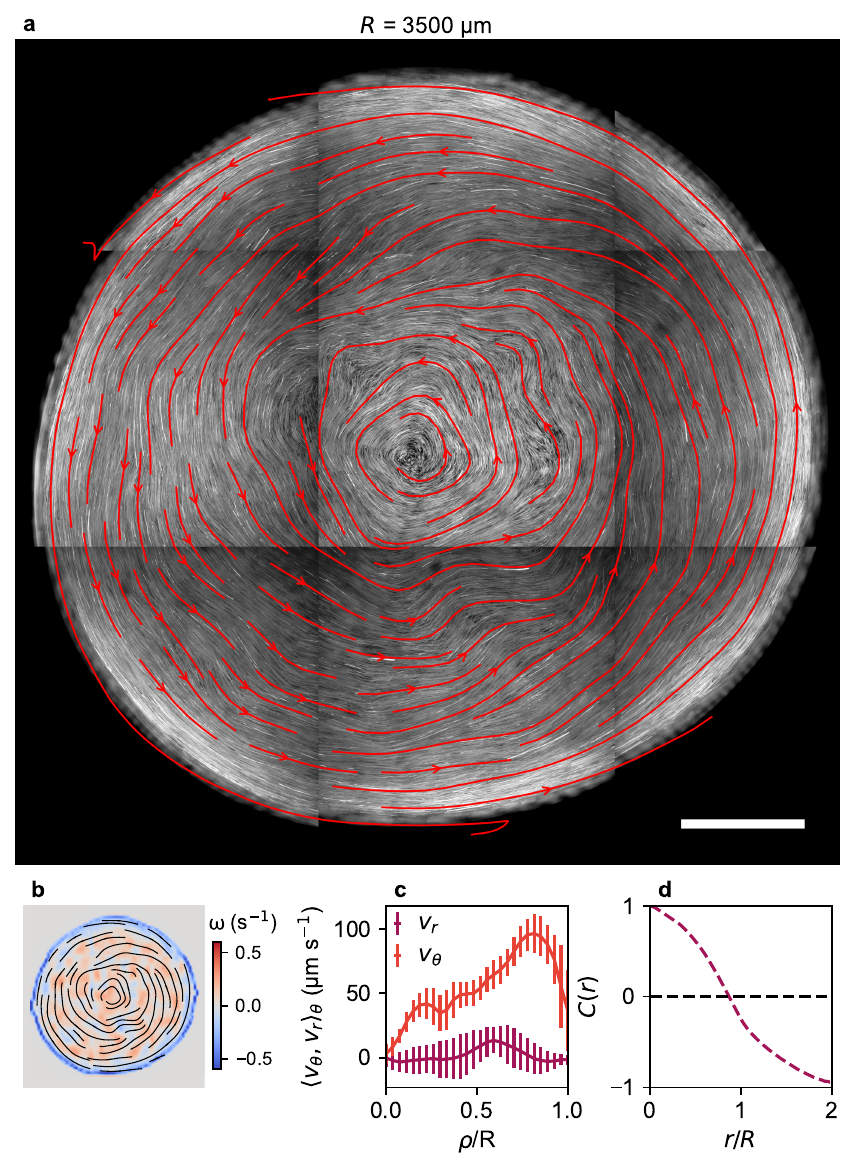}
\caption{ \textbf{SLV state for a slip boundary condition reaching $Re=0.26$.}  
\textbf{a} Cartography reconstruction of the flow in the sample. Radius $R=$~\qty{3500}{\micro\meter} and height $H=$~\qty{840}{\micro\meter}. Scale bar is \qty{1}{\milli\meter}. The black region represents air. 
\textbf{b} Vorticity field $\omega=|\nabla \times \bf{v}|$. 
\textbf{c} Components of the polar velocity profile. Purple is the radial component $v_r$, and orange is the angular component $v_\theta$. The root mean square of $v_\theta$ is $v_{RMS}=$~\qty{73}{\micro\meter\per\second} and the maximum of $v_\theta$ is $v_{max}=$~\qty{134}{\micro\meter\per\second}
\textbf{d} Spatial correlation function $C(r)$ calculated over the average flow. The crossover through zero occurs at $\lambda_0=$~\qty{3050}{\micro\meter}. Using $v_{RMS}$ and $R$ as characteristic scales gives $Re=0.26$, the maximum $Re$ we have observed in our experiments.
}
\label{fig:SLV_Re}
\end{figure}

\beginsupplement




\end{appendices}

\FloatBarrier


\bibliographystyle{unsrt}
%\bibliography{zotero_references}% common bib file
\bibliography{references}% temporary bib file

\end{document}